\documentclass[draft]{agujournal2019}
\usepackage{url} %this package should fix any errors with URLs in refs.
\usepackage{lineno}
\usepackage[inline]{trackchanges} %for better track changes. finalnew option will compile document with changes incorporated.
\usepackage{soul}
\usepackage{amsthm,amsmath,amssymb}
\usepackage{mathrsfs}
\usepackage{bm}
\usepackage{booktabs}
\usepackage{tabularx}
\draftfalse

\journalname{XXXXXX}

\begin{document}

\title{Improved Forecasts of Global Extreme Marine Heatwaves Through a Physics-guided Data-driven Approach}

\authors{Ruiqi Shu\affil{1}, Hao Wu\affil{2}, Yuan Gao\affil{1}, Fanghua Xu\affil{1}, Ruijian Gou\affil{3}, Xiaomeng Huang\affil{1}}

\affiliation{1}{Department of Earth System Science, Ministry of Education Key Laboratory for Earth System Modeling, Institute for Global Change Studies, Tsinghua University, Beijing, China}
\affiliation{2}{School of Computer Science and Technology, University of Science and Technology of China}
\affiliation{3}{Key Laboratory of Physical Oceanography and Frontiers Science Center for Deep Ocean Multispheres and Earth System, Ocean University of China, Qingdao, China.}

\correspondingauthor{Xiaomeng Huang}{hxm@tsinghua.edu.cn}

\begin{keypoints}
\item A purely data-driven model for global extreme marine heatwaves (MHWs) forecast is designed based on the physical nature of MHWs.
\item Motivated by ensemble forecasting techniques, two modules are designed to boost the predictability of the most extreme MHWs. 
\item We discuss the key factor of MHWs' small-scale variation through explainable AI methods.
\end{keypoints}

\begin{abstract}
The unusually warm sea surface temperature events known as marine heatwaves (MHWs) have a profound impact on marine ecosystems. Accurate prediction of extreme MHWs has significant scientific and financial worth. However, existing methods still have certain limitations, especially in the most extreme MHWs. In this study, to address these issues, based on the physical nature of MHWs, we created a novel deep learning neural network that is capable of accurate 10-day MHW forecasting. Our framework significantly improves the forecast ability of extreme MHWs through two specially designed modules inspired by numerical models: a coupler and a probabilistic data argumentation. The coupler simulates the driving effect of atmosphere on MHWs while the probabilistic data argumentation approaches significantly boost the forecast ability of extreme MHWs based on the idea of ensemble forecast. Compared with traditional numerical prediction, our framework has significantly higher accuracy and requires fewer computational resources. What's more, explainable AI methods show that wind forcing is the primary driver of MHW evolution and reveal its relation with air-sea heat exchange. Overall, our model provides a framework for understanding MHWs' driving processes and operational forecasts in the future.
\end{abstract}

\section*{Plain Language Summary}
Marine heatwaves (MHWs) are unusually warm seawater that occur on the surface of the global ocean. Accurate forecast of these MHWs is important for both science and finance. However, existing methods still have certain limitations, especially in the most extreme events. To tackle these problems, we use artificial intelligence (AI) methods to design a framework that can predict regional and global MHWs up to 10 days in advance. Our approach, which includes two specially designed modules, significantly improves the prediction of extreme MHWs. Compared to other existing methods, our approach performs better. We also find that in our model, wind forcing is the main driver of small-scale MHW changes and explains its relation with air-sea interactions. Overall, our model offers a new way to understand what causes MHWs and how to make better predictions for the future.
%%%%%%%%%%%%%%%%%%%%%%%%%%%%%%%%%%%%%%%%%%%%%%%
%
%  INTRODUCTION
%
%%%%%%%%%%%%%%%%%%%%%%%%%%%%%%%%%%%%%%%%%%%%%%%

\section{introduction}
Marine heatwaves (MHWs) represent anomalous warm seawater events that have a profound impact on marine ecosystems \cite{oliver2021marine, pearce2011marine}. For example, MHWs, especially those most extreme ones, can cause coral bleaching \cite{hughes2017global, hughes2018spatial} and widespread mortality of marine organisms \cite{garrabou2009mass,thomson2015extreme}. As a result, an accurate forecast of extreme MHWs has far-reaching scientific and economic value. For example, sub-seasonal MHWs forecasting can help sea food production and management planning, such as feed cycles, at 1-7 day timescales while seasonal forecasting can further support proactive decision-making for the blue economy \cite{hobday2016seasonal, malick2020environmentally, mills2017forecasting, payne2022skilful}. In this study, we will mainly focus on the sub-seasonal forecast.\par

Traditional forecasting methods for MHWs are often based on numerical models in which the primitive equations of the ocean are used to simulate the variation of MHWs. Such forecasts generally fall into two categories. The first is seasonal forecasting, the main purpose of which is to give a general pattern of onset, duration, and intensification of MHWs over a long timescale \cite{jacox2022global,brodie2023ecological}. The other type is sub-seasonal forecasting, which is mainly designed to measure the small-scale variations of MHWs. Such forecast is more accurate than seasonal forecasts, but the forecast lead time is much shorter \cite{benthuysen2021subseasonal, yu2024assessing}. Although physics-based numerical models achieve a satisfying forecast of MHWs, they require large amounts of computer resources \cite{xiong2023ai, wang2024xihe} and show limited ability in forecasting the extreme MHW events \cite{jacox2022global}. \par

Currently, with the development of artificial intelligence, deep learning techniques have recently been applied to global weather and ocean forecasts with impressive forecasting abilities. \cite{bi2022pangu,chen2023fengwu,chen2023fuxi,kurth2023fourcastnet,xiong2023ai,wang2024xihe}. When it comes to the MHWs and their associated extreme sea surface temperature (SST), several approaches have been made:

Paradigm 1. End-to-end marine heatwaves forecast. The core idea is to model the evolution of SST anomaly (SSTA) through machine learning techniques or neural networks combing with some physical prior (such as Rossby normal mode in the South China Sea \cite{lin2023rossby})\cite{giamalaki2022assessing, sun2023artificial, shao2020mid}. \par
%These pioneering studies have revealed the potential of AI methods in marine heatwave forecasting, but there are still some limitations, such as the lack of full consideration of the atmosphere impact and the limited ability to capture extreme heatwave events. 
Paradigm 2. Large AI-driven ocean models. Recently, with the surprising success of AI-driven weather models \cite{bi2022pangu,chen2023fengwu,chen2023fuxi,kurth2023fourcastnet,lam2023learning}, some AI-driven ocean models capable of predicting essential ocean variables (including SST) have been proposed \cite{wang2024xihe,xiong2023ai}. \par
%However, these large-AI driven ocean models require the future atmospheric field as an additional input for autoregressive forecasting, which will encounter difficulties in operational forecasting. 
Paradigm 3. Hybrid model based on numerical models and data-driven methods. These methods are usually based on the results of physical predictions, which use neural networks to achieve the effect of debiasing \cite{sun2024deep}. \par
%However, these models still rely on the results of numerical predictions and a large amount of forecast data.
However, the forecast abilities of these approaches are still limited, especially in those most extreme MHW events \cite{giamalaki2022assessing}. There are two main reasons for this shortcoming of data-driven models: First, some models do not fully consider the impact of atmospheric forcing, which is often one of the most important trigger factors for the occurrence of extreme MHWs \cite{holbrook2019global}. Second, since the aim of a deep learning model is searching for the global optimal solution, the model is more inclined to generate a conservative prediction. This is reflected in the forecast results, which are overly smoothed SST fields and fewer extreme MHW events. \par

%there is still some critical problem concerning the data-driven extreme MHWs prediction: First, these data-driven approaches' forecasting abilities are limited since the driving effect of the atmosphere on MHWs is not fully taken into account. Second, most of the traditional spatiotemporal forecasting techniques based on deep learning are deterministic forecasts. Since the training of a deep learning model is searching for the global optimal solution, the model is more inclined to generate a conservative prediction. This is reflected in the forecasting of marine heatwaves, which are overly smoothed SST fields and fewer extreme MHW events \cite{giamalaki2022assessing}.\par

By addressing the two shortcomings of the data-driven methods mentioned above, in this study, we designed a global data-driven framework consists of two modules that significantly improves the forecasts of extreme MHWs inspired by numerical models. To be specific, first, we designed a coupler to fully incorporate the role of atmospheric in driving extreme MHWs together with the state-of-the-art deep learning techniques. Second, based on the idea of ensemble forecasting in numerical forecasting, we also designed a probabilistic data augmentation method to improve the model's forecast ability for extreme MHW events. Compared to previous data-driven forecasting models, our framework has demonstrated significant performance improvements that can reliably generate 10 day forecasts of global extreme MHWs. Our framework shows a potential competitiveness to the numerical models, and the time required to complete a forecast is orders of magnitude smaller. What's more, through explainable methods, we found that the surface wind speed is the key driving factor of subseasonal MHWs variations, which impacts SSTA through latent heat flux. 
%%%%%%%%%%%%%%%%%%%%%%%%%%%%%%%%%%%%%%%%%%%%%%%
%
%  Materials and Methods
%
%%%%%%%%%%%%%%%%%%%%%%%%%%%%%%%%%%%%%%%%%%%%%%%

\section{Materials and Methods}
\subsection{Data}
In this study, we utilize two sets of reanalysis datasets of the ocean surface variables: the daily 1/4° eddy-permitting dataset and the 1/12° eddy-resolving dataset  \cite{lellouche2018copernicus}. The former is used to generate global eddy-permitting forecast, while the latter is for regional eddy-resolving forecast. Based on the different SST datasets, we conduct two experiments: one is the global 1/4° eddy-permitting MHW forecast, and the other is the North Pacific regional 1/12° eddy-resolving MHW forecast. \par
We utilize the \citeA{hobday2016seasonal} definition of MHW categories here. More precisely, any interval in which the SST anomalies surpass the 90th percentile for a minimum of five days in a row is classified as an MHW event. Subsequently, the MHWs are divided into three severity categories: moderate, strong, and severe/extreme. These categories are determined by the 90th, 92.5th, and 95th percentiles of the SSTA, respectively \cite{hobday2016seasonal}. \par
The atmospheric variables are obtained from ECWMF Reanalysis v5 (ERA5) \cite{hersbach2020era5}. Based on the physical process, we select three variables: 10m wind components ($U_{{\rm 10}}$, $V_{{\rm 10}}$), 2m temperature ($T_{{\rm 2m}}$), and four surface heat flux variables (latent and sensible heat flux $Q_{{\rm lat}}, Q_{{\rm sens}}$, shortwave and longwave radiation $Q_{{\rm SW}}, Q_{{\rm LW}}$). The detailed description of the data can be found in Table 1: 

\begin{table*}[!htbp]
    \centering
    \caption{Detailed description of the dataset}
    \begin{tabularx}{\textwidth}{>{\centering\arraybackslash}X>{\centering\arraybackslash}X>{\centering\arraybackslash}X>{\centering\arraybackslash}X}
        \toprule
        \textbf{Variable Name} & \textbf{Resolution} & \textbf{Spatial Range} & \textbf{Temporal Range} \\
        \midrule
        $U_{{\rm 10}}$ & 1/4° & 90°S-90°N, 180°E-180°W & daily snapshot at 12:00 UTC, 1993-2021 \\
        $V_{{\rm 10}}$ & 1/4° & 90°S-90°N, 180°E-180°W & daily snapshot at 12:00 UTC, 1993-2021 \\
        $T_{{\rm 2m}}$ & 1/4° & 90°S-90°N, 180°E-180°W & daily snapshot at 12:00 UTC, 1993-2021 \\
        $Q_{{\rm lat}}$ & 1/4° & 90°S-90°N, 180°E-180°W & daily mean, 1993-2021 \\
        $Q_{{\rm sens}}$ & 1/4° & 90°S-90°N, 180°E-180°W & daily mean, 1993-2021 \\
        $Q_{{\rm SW}}$ & 1/4° & 90°S-90°N, 180°E-180°W & daily mean, 1993-2021 \\
        $Q_{{\rm LW}}$ & 1/4° & 90°S-90°N, 180°E-180°W & daily mean, 1993-2021 \\
        SSTA (eddy-permitting) & 1/4° & 80°S-90°N, 180°E-180°W & daily mean, 1993-2021 \\
        SSTA (eddy-resolving) & 1/12° & 22.5°N-47.5°N, 140°E-120°W & daily mean, 1993-2021 \\
        \bottomrule
    \end{tabularx}
    \label{tab:dataset}
\end{table*}

Additionally, we are going to demonstrate the competitiveness of our forecasting framework against conventional numerical forecasting using GOFS 3.1 global ocean variable forecasting data, which is one of the leading numerical forecasts based on a 1/12° HYCOM model and NCODA system \cite{helber2013validation,cummings2005operational,cummings2013variational}. To facilitate comparison, we downsample the 1/12° resolution of the original data to 1/4°.  In this study, we use data from 1993-2021 for a total of 29 years, with 1993-2017 as the training/validation set and 2018-2021 as the test set. All input variables mentioned above are normalized before they are fed into our model:
\begin{equation}
\begin{aligned}
    \overline{\mathcal{X}} = \frac{\mathcal{X}-{\rm mean(\mathcal{X})}}{{\rm std(\mathcal{X})}} 
\label{normalized}
\end{aligned}
\end{equation}
where the mean value and standard derivation are calculated over the training set.

\subsection{Deterministic Global MHWs Forecast Framework}
As shown in Figure 1, our model can be separated into two modules: the deterministic forecast module (Figure 1a) and the probabilistic data argumentation module (Figure 1b). The problem of marine heatwave forecast is essentially predicting the future state of mixed layer SSTA $\big\langle T_{t} \big\rangle^{\prime}$ based on the current state $\big\langle T_{0} \big\rangle^{\prime}$, here the brackets denote the average over the mixed layer and the prime is the deviation from the climatology mean. To be specific, we seek to approximate the evolution operator $\phi$:
\begin{equation}
\begin{aligned}
    \big\langle T_{t} \big\rangle^{\prime} = \phi(\big\langle T_{0} \big\rangle^{\prime}, A_{0:t})
\label{evolution operator}
\end{aligned}
\end{equation}\par
In all, our goal is to use a neural network $\phi_{\theta}$ to approximate $\phi$ based on the existing data. The governing equation of $\big\langle T \big\rangle^{\prime}$ can be written as \cite{bian2023oceanic}:
\begin{equation}
\begin{aligned}
    \big\langle\frac{\partial T}{\partial t}\big \rangle^{\prime} = \big\langle -\nabla(\textbf{u}T) \big \rangle^{\prime} + \frac{Q_{{\rm SW}}^{\prime}+Q_{{\rm LW}}^{\prime}+Q_{{\rm sens}}^{\prime}+Q_{{\rm lat}}^{\prime}}{\rho C_{p}h} + \big\langle {\rm MIX} \big \rangle^{\prime}
\label{heat budget}
\end{aligned}
\end{equation}
where $\textbf{u}$ represents the horizontal velocity, and $Q_{{\rm SW}}, Q_{{\rm LW}}, Q_{{\rm sens}}, Q_{{\rm lat}}$ is the shortwave radiation, longwave thermal radiation, sensible heat flux, and latent heat flux respectively. Integrate (\ref{heat budget}) we get
\begin{equation}
\begin{aligned}
    \big\langle T_{t }\big\rangle^{\prime} - \big\langle T_{0} \big\rangle ^{\prime}&= \int_{0}^{t} \big\langle -\nabla(\textbf{u}T) \big \rangle^{\prime}  + \frac{Q_{{\rm SW}}^{\prime}+Q_{{\rm LW}}^{\prime}+Q_{{\rm sens}}^{\prime}+Q_{{\rm lat}}^{\prime}}{\rho C_{p}h} dt + {\rm RES}\\
\label{heat budget integrated}
\end{aligned}
\end{equation}
%Here $\textbf{u} = \textbf{u}_{E} + \textbf{u}_{O}$ follows a similar decomposition as in \citeA{bian2024scale}.  \citeA{bian2024scale} points out that the effect of the atmosphere boundary condition is mainly reflected in the heat flux, so the first term in (\ref{heat budget integrated}) is negligible. 
Guided by this physical nature of SSTA, we first use train a neural network (called a `coupler') to approximate heat flux $Q^{\prime} = \phi_{\theta}^{(1)}(A_{0:t}, \big\langle T_{0}\big\rangle^{\prime})$. Other effects, including advection and mixing effects, are approximated by another neural network $\big\langle T_{t} \big\rangle^{\prime} = \phi_{\theta}^{(2)}(Q^{\prime}, \big\langle T_{0}\big\rangle^{\prime})$. In all, our framework can be expressed as:
\begin{equation}
\begin{aligned}
    \big\langle T_{t} \big\rangle^{\prime} = \phi_{\theta}^{(2)}(\phi_{\theta}^{(1)}(A_{0:t}, \big\langle T_{0}\big\rangle^{\prime}), \big\langle T_{0}\big\rangle^{\prime})
\label{framework}
\end{aligned}
\end{equation}
In the training phase, the future atmosphere condition $A_{0:t}$ is the ground truth data from ERA5 while in the operational forecast, it's replaced by the Pangu-weather's forecast initialized by ERA `$t=0$' analysis field \cite{bi2022pangu}. Furthermore, the AI-based Pangu-weather model only requires atmospheric variables as inputs, ensuring our framework's self-consistency. \par

The specific structure of $\phi_{\theta}^{(1)}$ and $\phi_{\theta}^{(2)}$ is flexible and can be replaced with a variety of neural networks such as U-Net, SimVP, ConvLSTM, PastNet \cite{ronneberger2015u,gao2022simvp,shi2015convolutional,wu2023pastnet}. During the training and inference phase, the input variables is arrange into a 4-dimensional tensor $\mathcal{X} \in {\mathbb R}^{T*C*H*W}$, where $T=10$ is the forecast lead time, $C$ is the number of input variables, $(H,W)=(1440,720)$ (for eddy-permitting experiments) $(1200,300)$ (for eddy-resolving experiments) represents the horizontal resolution. 

\subsection{Probabilistic Approach for Most Extreme MHWs}

%Text here ===>>>
As pointed out in previous sections, overly smoothed forecast results will forbid AI-based model's forecast ability on the most extreme MHW events. A key factor is that samples on extreme heatwaves are actually very scarce (less than $10 \%$ by the definition of MHWs). Therefore, we need to generate more training samples with more extreme MHWs that conform to physical laws based on existing datasets, so as to improve the forecast ability of AI models in extreme events (i.e., data argumentation). \par
In numerical models, more forecast results can be obtained by ensemble forecasts. To be specific, under more realistic circumstances, small perturbations will be added to the initial value $\big\langle T_{0}\big\rangle^{\prime}$ and boundary condition $A_{0:t}$ due to the uncertainty in observation and subgrid scale signals. Thus, in actuality, $\big\langle T_{t}\big\rangle^{\prime}$ follows a particular probabilistic distribution $\big\langle T_{t}\big\rangle^{\prime}\sim p(\big\langle T_{t}\big\rangle^{\prime}|\big\langle T_{0}\big\rangle^{\prime},A_{0:t})$. In traditional numerical ensemble forecast, by adding perturbations that conform to certain patterns (for example, random noise or singular vectors), one can sample from this conditional distribution and boost the forecast ability. \par

Nevertheless, due to the fact that current AI-based models lack the chaotic effect \cite{selz2023can, chen2024machine}, deep learning-based techniques cannot simply adopt the paradigm of adding initial perturbation. As a result, inspired by \citeA{wu2024beamvq}, rather than simply sample from the conditional distribution $ p$, we utilized a vector quantised-variational autoencoder (VQVAE) to learn this distribution directly based on Bayesian theory \cite{van2017neural}. Specifically, our probabilistic approach consists of the following steps:\par

\textbf{Step 1 : Pretrain the Deterministic Model}. First, following section 2.2, we train a deterministic forecast model $\phi_{\theta} = \phi_{\theta}^{(1)} \circ \phi_{\theta}^{(2)} $ based on the existing training dataset $\{ \mathcal{X}_{i}\}_{i=1}^{N}$ and model output $\{\mathcal{T}_{i} = \phi_{\theta} (\mathcal{X}_{i})\}_{i=1}^{N}$. However, as shown in Figure 4, those output is always overly smoothed due to the intrinsic deficiency of AI-based methods.\par

\textbf{Step 2 : Learning the Conditional Distribution}. Second, we use VQVAE to learn conditional distribution $p(\big\langle T_{t}\big\rangle^{\prime}|\big\langle T_{0}\big\rangle^{\prime},A_{0:t})$ (i.e., the distribution of SSTA $\big\langle T_{t}\big\rangle^{\prime}$ in the next 10 days, given the initial and boundary conditions). VQVAE model (abbreviated as $\Psi$) consists of two nerual networks: an encoder $q_{\phi}$ and a decoder $p_{\theta}$ (i.e., $\Psi=p_{\theta} \circ q_{\phi}$). In encoder, we want to transform the input $\big\langle T_{t}\big\rangle^{\prime}$ into a discrete latent space $\mathcal{Z}$. The latent representation is then fed to the decoder to reconstruct $\big\langle T_{t}\big\rangle^{\prime}$.  As a result, Once the VQVAE is trained, we can simply sample in the $z \in \mathcal{Z}$ and use the decoder to acquire the conditional distribution $p$: $q_{\phi}(z) \sim p(\big\langle T_{t}\big\rangle^{\prime}|\big\langle T_{0}\big\rangle^{\prime},A_{0:t})$. \par
Mathematically, the role of encoder and decoder can be written as conditional distributions $q_{\phi}(z|\big\langle T_{t}\big\rangle^{\prime},\big\langle T_{0}\big\rangle^{\prime},A_{0:t})$ and $p_{\theta}(\big\langle T_{t}\big\rangle^{\prime}|z, \big\langle T_{0}\big\rangle^{\prime},A_{0:t})$. The goal of our model is to learn the distribution of $p$, that is, to minimize the 'distance' between $q_{\phi}$ , $p_{\theta}$ and $p$. The former can be realized through minimizing the Kullback-Leibler Divergence (K-L Divergence,  here we denote $X=(\big\langle T_{t}\big\rangle^{\prime},\big\langle T_{0}\big\rangle^{\prime},A_{0:t})$ for simplicity)
\begin{equation}
\begin{aligned}
\int q_{\phi}(z|X) {\rm log} (\frac{q_{\phi}(z|X)}{p(z|X))})dz
\\ \triangleq
D_{KL}(q_{\phi}(z|X), p(z|X))
\label{kl_div}
\end{aligned}
\end{equation}
while the latter is equivalent to the maximize the maximum likelihood estimation ${\rm log} p_{\theta}$. As a result, the model is optimized through evidence lower bound  (ELBO) $\mathcal{L}(q_{\phi}, p_{\theta})$ of $p_{\theta}$ and $q_{\phi}$ to ensure the distribution of the output of $\Psi$ is as close as $p(\big\langle T_{t}\big\rangle^{\prime}|\big\langle T_{0}\big\rangle^{\prime},A_{0:t})$:

\begin{equation}
\begin{aligned}
\mathcal{L} 
&= -D_{KL}(q_{\phi}(z|X), p(z|X)) + {\rm log} p_{\theta}(\big\langle T_{t}\big\rangle^{\prime}|\big\langle T_{0}\big\rangle^{\prime},A_{0:t})\\
&=-\int q_{\phi}(z|X) {\rm log} (\frac{q_{\phi}(z|X)}{p(z|X))})dz + \int q_{\phi}(z|X) {\rm log} p_{\theta}(\big\langle T_{t}\big\rangle^{\prime}|\big\langle T_{0}\big\rangle^{\prime},A_{0:t}) dz\\
&= \int q_{\phi}(z|X) {\rm log} (\frac{p_{\theta}(\big\langle T_{t}\big\rangle^{\prime}|\big\langle T_{0}\big\rangle^{\prime},A_{0:t})p(z|X)}{q_{\phi}(z|X)})dz\\
&= \int q_{\phi}(z|X) {\rm log} (\frac{p(z|\big\langle T_{0}\big\rangle^{\prime},A_{0:t})p_{\theta}(\big\langle T_{t}\big\rangle^{\prime}|z, \big\langle T_{0}\big\rangle^{\prime},A_{0:t})}{q_{\phi}(z|X)})dz\\
&=-D_{KL}(q_{\phi}(z|X), p(z|\big\langle T_{0}\big\rangle^{\prime},A_{0:t})) + \mathbb{E}_{z \sim q_{\phi}(z|X)}[{\rm log} p_{\theta}(\big\langle T_{t}\big\rangle^{\prime}|z, \big\langle T_{0}\big\rangle^{\prime},A_{0:t})]
\end{aligned}
\label{vaeloss}
\end{equation}

Where $D_{KL}$ is the K-L Divergence and $\mathbb{E}_{q_{\phi}}[{\rm log} p_{\theta}(\big\langle T_{t}\big\rangle^{\prime}|z, \big\langle T_{0}\big\rangle^{\prime},A_{0:t})]$ is the expectation of ${\rm log} p_{\theta}(\big\langle T_{t}\big\rangle^{\prime}|z, \big\langle T_{0}\big\rangle^{\prime},A_{0:t})$ under distribution $q_{\phi}$. Since $q_{\phi}(z|\big\langle T_{t}\big\rangle^{\prime},\big\langle T_{0}\big\rangle^{\prime},A_{0:t})$ in fact follows a discrete distribution, direct derivation shows that $\mathcal{L}$ is equivalent to the mean square error (MSE) loss (the K-L divergence term vanishes while the $\mathbb{E}_{z \sim q_{\phi}(z|X)}[{\rm log} p_{\theta}(\big\langle T_{t}\big\rangle^{\prime}|z, \big\langle T_{0}\big\rangle^{\prime},A_{0:t})]$ is equivalent to MSE loss): 
\begin{equation}
\mathcal{L} = (\Psi(\big\langle T_{t}\big\rangle^{\prime})-\big\langle T_{t}\big\rangle^{\prime})^2
\label{vaeloss-simplified}
\end{equation}
In summary, based on the MSE loss, we can train the model to learn a more realistic distribution of  $p$.\par

\textbf{Step 3 : Data Augmentation Based on Forecast Ensembles}. 
In this step, we are going to generate forecast ensembles from the learned distribution $p(\big\langle T_{t}\big\rangle^{\prime}|\big\langle T_{0}\big\rangle^{\prime},A_{0:t})$. To be specific, given the trained VQVAE model $\Psi=p_{\theta} \circ q_{\phi}$, $\{ \phi_{\theta} (\mathcal{X}_{i})\}_{i=1}^{N}$ in step 1 with a Gaussian noise $z$ is input into encoder $q_{\phi}$ in order to generate latent representation $\{ z_{i}\}_{i=1}^{N}$. Then, we perturb these latent representations by finding the top-$k$ closest for each input sample (in this study, we choose $k=10$): 
\begin{equation}
z_{i}^{(1)},z_{i}^{(2)},...,z_{i}^{(k)} = {\rm min}_{z \in \mathcal{Z}}||z-q_{\phi}(\big\langle T_{t}\big\rangle^{\prime})||
\label{top-k}
\end{equation}
where $||\bullet||$ is the Euclidean distance in the latent space. Those latent vectors is then fed into the decoder $p_{\theta}$ to generate the forecast ensembles $\mathcal{T}_1^{i}, \mathcal{T}_2^{i}, \ldots,\mathcal{T}_k^{i}$:

\begin{equation}
    \{\mathcal{T}_1^{i}, \mathcal{T}_2^{i}, \ldots,\mathcal{T}_k^{i}\} = p_{\theta}(z_{i}^{(1)},z_{i}^{(2)},...,z_{i}^{(k)})=\psi(\phi_{\theta} (\mathcal{X}_{i}), z) \sim p(\big\langle T_{t}\big\rangle^{\prime}|\big\langle T_{0}\big\rangle^{\prime},A_{0:t})
\end{equation}
Since during the second step the trained decoder $p_{\theta}$ can already transform the latent representation $z \in \mathcal{Z}$ into the conditional distribution, those output forecast ensembles also follows the distribution $p(\big\langle T_{t}\big\rangle^{\prime}|\big\langle T_{0}\big\rangle^{\prime},A_{0:t})$ (i.e., more realistic SSTA fields under the initial and boundary conditions). In other words, by perturbing the latent representations, we acquire the forecast ensembles based on the deterministic forecast results.\par

\textbf{Step 4 : Iterative Training Based on Data argumentation}. 
Building upon the previous steps, in this step, we enhance the deterministic forecast model $\phi_{\theta}$ by incorporating high-quality samples generated from the learned conditional distribution $p$. Specifically, from the forecast ensembles ${\mathcal{T}_1^{i}, \mathcal{T}_2^{i}, \ldots, \mathcal{T}_k^{i}}$ generated in Step 3 for each input $\mathcal{X}_i$, we select a subset of $m$ high-quality samples ${\hat{\mathcal{T}}_1^{i}, \hat{\mathcal{T}}_2^{i}, \ldots, \hat{\mathcal{T}}_m^{i}}$ based on their proximity to the ground truth $\mathcal{T}_i^{\text{true}}$. The selection criterion is defined using the $L^2$ norm between the generated samples and the ground truth:
\begin{equation}
\hat{\mathcal{T}}_j^i=\underset{\mathcal{T}_j^i \in\left\{\mathcal{T}_1^i, \ldots, \mathcal{T}_k^i\right\}}{\arg \min }\left\| \mathcal{T}_j^i-\mathcal{T}_i^{\text {true }}\right\|_2^2, \quad j=1, \ldots, m .
\label{eq:select_high_quality} 
\end{equation}
The formula above is calculated for all grid points where MHW events occur. These selected high-quality samples are then combined with the original dataset to form an augmented training set:
\begin{equation} 
\mathcal{D}_{\text{aug}} = \left\{ (\mathcal{X}_i, \mathcal{T}_i^{\text{true}}) \right\}_{i=1}^{N} \cup \left\{ (\mathcal{X}_i, \hat{\mathcal{T}}_j^{i}) \right\}_{i=1}^{N}, \quad j = 1, \ldots, m. 
\label{eq:augmented_dataset}
\end{equation}
We then retrain the deterministic model $\phi_{\theta}$ on the augmented dataset $\mathcal{D}_{\text{aug}}$ to improve its forecasting performance. The loss function for retraining is defined as:
\begin{equation}
\mathcal{L}_{\text {aug }}=\frac{1}{N+N m} \sum_{i=1}^N\left[\left\|\phi_\theta\left(\mathcal{X}_i\right)-\mathcal{T}_i^{\mathrm{true}}\right\|_2^2+\sum_{j=1}^m\left\|\phi_\theta\left(\mathcal{X}_i\right)-\hat{\mathcal{T}}_j^i\right\|_2^2\right]
\end{equation}
This retraining process leverages both the original ground truth and the high-quality generated samples to refine the model. By incorporating the selected samples that closely resemble the ground truth, the model learns to better capture the variability and extremes present in the data. \par

In summary, this approach not only enhances the deterministic model's performance but also mitigates the over-smoothing issue observed in AI-based forecasting methods.

\subsection{Evaluation Metrics}
In this study, to evaluate the performance of our model, we use three metrics: Root Mean Square Error (RMSE), Critical Success Index (CSI) and Symmetric Extremal Dependence Index (SEDI). RMSE represents the overall performance of our SST anomaly forecast, which is calculated as: 
\begin{equation}
\begin{aligned}
{\rm RMSE}(t) = \sqrt{\frac{1}{H \times W}\sum_{i,j}(T_{i,j}(t)-\tilde{T}_{i,j}(t))^2}
\label{RMSE}
\end{aligned}
\end{equation}
Where $T_{i,j}(t)$ and $\tilde{T}_{i,j}(t)$ is the ground-truth and prediction of SST anomaly at lead time $t$. The subscript $(i,j)$ indicates the data at grid point $(i,j)$ while H and W follows the same definition in section 2.1. CSI and SEDI are the metrics that evaluate the forecast ability of extreme events and can be expressed as:
\begin{equation}
\begin{aligned}
{\rm CSI}(t) = \frac{{\rm TP}}{{\rm TP}+{\rm FP}+{\rm FN}}
\label{CSI}
\end{aligned}
\end{equation}
and
\begin{equation}
\begin{aligned}
{\rm SEDI}(t) = \frac{{\rm log}(F)-{\rm log}(H)-{\rm log}(1-F)+{\rm log}(1-H)}{{\rm log}(F)+{\rm log}(H)+{\rm log}(1-F)+{\rm log}(1-H)}
\label{SEDI}
\end{aligned}
\end{equation}
Where TP (True Positive) denotes the number of cases that a MHWs event is accurately predicted. FP (False Positive), FN (False Negative), and TN (True Positive) follow a similar definition. $F=\frac{FP}{FP+TP}$ is the false alarm rate while $H=\frac{TP}{TP+FN}$ is the hit rate. \par

In this study, we also utilize explainable AI (XAI) method to evalute the relative importance of each input of our model $\phi_{\theta}$. To be specific, by removing the $i^{\rm{th}}$ variables from our input, we can quantitatively measure the (unsigned) contribution of each variable on the heat flux (called `contribution map') at grid point $(m,n)$, which is calculated as:
\begin{equation}
\begin{aligned}
\Delta C^{ns}_{j} (m,n) =\sqrt{\frac{\sum_{i \in Y}^{N}(\phi_{\theta}^{(1)}(\mathcal{X}_{i}(m,n)) - \phi_{\theta}^{(1)}(\mathcal{X}_{i}^{j}(m,n)))^2}{|Y|}}
\label{importance_ns}
\end{aligned}
\end{equation}
The signed contribution can be calculate as:
\begin{equation}
\begin{aligned}
\Delta C^{s}_{j} (m,n) =\frac{\sum_{i \in Y}^{N}\phi_{\theta}^{(1)}(\mathcal{X}_{i}(m,n)) - \phi_{\theta}^{(1)}(\mathcal{X}_{i}^{j}(m,n))}{|Y|}
\label{importance_s}
\end{aligned}
\end{equation}
Where $\{ \mathcal{X}_{i}\}_{i=1}^{N}$ is the full input variables while $\{ \mathcal{X}_{i}^{j}\}_{i=1}^{N}$ is the modified input with the $j^{\rm{th}}$ element replaced with zero. $Y$ represents the all MHW events and $|Y|$ is the number of MHW events. Similar approaches can be found at \cite{shin2022application, shin2024deep, lyu2023improving}. \par
%%%%%%%%%%%%%%%%%%%%%%%%%%%%%%%%%%%%%%%%%%%%%%%
%
%  RESULTS
%
%%%%%%%%%%%%%%%%%%%%%%%%%%%%%%%%%%%%%%%%%%%%%%%
\section{Results}
\subsection{A Global Assessment of Our Frameworks' Performance}
Figure 2 illustrates the performance of our framework in the global 1/4° dataset. Overall, our framework's forecasting ability for $\big\langle T\big\rangle^{\prime}$ exceeds existing data-driven models \cite{giamalaki2022assessing, sun2023artificial}, and the mean RMSE of 10-day forecast can be reduced by $30\%$ to $40\%$. In addition, our framework has shown to be quite competitive compared to numerical models. Since there is a certain deviation of SST in different sources of observation or analysis data, in this study, we use GOFS's analysis field as the input initial condition of the model to ensure the fairness of the comparison. However, despite the unfavorable conditions for our framework, our model's 6-day forecast RMSE is still $20\%$ to $30\%$ lower than that of the numerical model. What's more, as shown in Figure 2a, the boundary condition $A_{0:t}$ plays an important role in the forecasting of $\big\langle T\big\rangle^{\prime}$. Under a realistic operational forecast condition, the $A_{0:t}$ is replaced by the predicition of Pangu-weather, which will impair the forecast ability of $\big\langle T\big\rangle^{\prime}$ (with the RMSE increasing about $20\%$, see the solid and dash blue line in Figure 2a). In other words, as the lead time marches on, the accuracy of the estimation of the atmospheric condition $A_{0:t}$ has an increasing impact on our framework. \par

What's more, when it comes to the extreme SSTA events (i.e., marine heatwaves), our model has also shown good forecasting capabilities. Figure 2b-2c shows the global average for CSI and SEDI indicators. Comparing to the existing data-driven methods, our model also significantly improves the ability to capture such MHWs. Coastal ecosystems are the richest and most complex in the ocean \cite{costello2017marine, halpern2008global}. In this study, we also evaluate the forecast skills of coastal MHWs against the numerical model over 11 coastal regions following \citeA{marin2021slower} (Figure 2d). There is a $20\%$ to $50\%$ improvement in CSI of each region, especially along the coastline of Somalia (SOM) where numerical model can only capture a small portion of MHW events. This provided a potential application of our model to coastal MHWs forecast.\par

We also evaluate our framework's forecast ability in well-known, long-persisting extreme MHWs events. For example, within the Central North Pacific region, strong MHWs are frequently observed during 2019-2021 summer (Abbreviated as CNP-MHW in the following paragraphs). Here we take 2021 CNP-MHW as an example, as shown in Figure 2e-2f. More examples can be found in Figure S2. Based on the heat budget equations (\ref{heat budget}), the `surface effects', including the advection (ADV) and heat flux (FLUX) term, can only explain about $30 \%$ of the total changes of SSTA during the intensifying and decaying phase of CNP-MHW as in Figure 2e. In fact, based on Argo profiling data, \citeA{nishihira2024record} found that during 2021 Summer, the Central Mode Water (CMW) decreased extremely, leading to less cooling heat flux associated with the entrainment of subsurface waters into mixed layer, which contributed to form the CNP-MHW. This means that after receiving the FLUX term from coupler $\phi_{\theta}^{(1)}$, our main model $\phi_{\theta}^{(2)}$ can parameterize the subsurface mixed layer entrainment process based on the surface data. Overall, our approach has the potential capacity to replicate the subsurface process from surface data, demonstrating our model's robustness under more complex MHW patterns.

In addition to the high accuracy of MHWs forecasting, our framework also significantly improves the efficiency. To be specific, our model only takes 0.2 seconds to complete a 10-day global MHW forecast on a single A100 GPU, which is several orders of magnitude faster than numerical forecast systems \cite{bi2022pangu,wang2024xihe}. \par

\subsection{Forecast Ability of the Extreme MHWs}
Unlike numerical models or large-AI driven ocean model that treats the SST as a primitive variable, we are more concerned with capturing the extremes of the SST. In this study, we apply the probabilistic method in section 2.3 to our deterministic forecasting, which effectively improves the overall forecasting ability of the framework and extreme MHWs. To be more explicit, as demonstrated in Figures 3a-3b, after incorporating the probabilistic module, all of the metrics steadily improve. Even in the most intense MHWs incidents (99th percentile), the CSI still increases by approximately $15\%$.\par
Here we used the CNP-MHW in 2020 (as mentioned in section 3.1) to illustrate how our forecast ensembles improves the forecast ability of extreme events (we don't take 2021 CNP-MHW mainly because the backbone module can already generate satisfying result). Figure 3d shows a 10-day forecast of the CNP-MHW event initialized at 2020-08-14 and ending at 2020-08-24. The level of CNP-MHW gradually rose from moderate to extreme, broke the $99\%$ threshold on 2020-08-19, and then steadily declined. Our deterministic forecasting framework can also capture the formation of CNP-MHW, but it is significantly underestimated in intensity. Using the probabilistic model mentioned in Section 2.3, we sampled $k=10$ forecast ensembles from the conditional distribution $p(\big\langle T_{t}\big\rangle^{\prime}|\big\langle T_{0}\big\rangle^{\prime},A_{0:t})$ and added them to the training samples. It can be found that most of the 10 forecast ensembles capture extreme temperature enhancements, and their intensity is higher than that of the deterministic model, which is closer to the real situation. As a result, the new forecasting model trained on the enhanced forecast ensembles dataset can accurately capture the CP-MHW enhancement process and improve the prediction ability of extreme heat waves. A similar analysis is performed on the peak value of $\big\langle T_{t}\big\rangle^{\prime}$, which leads to a similar conclusion. \par
Adding those new forecast ensembles allows the dataset to be more widely distributed, allowing our forecast model to capture out-of-distribution MHW events, which can also be verified by Figure 3c. By calculating the probability density function (PDF) of $\big\langle T_{t}\big\rangle^{\prime}$, we find that the new probabilistic forecast model can lead to more extreme MHW events than conservative estimates based on deterministic forecast. In another word, our probabilistic-based data argumentation serves as a distribution shift that pulls the conservative and over-smoothing forecast back to more extreme and rare cases. \par 

Another question is about our framework's performance in the eddy-resolving regime. As pointed out by \citeA{bian2023oceanic,bian2024scale}, as the resolution increases, more small-scale MHWs will occur with life cycles of about 5-10 days. These MHWs are always small at spatial scale (around the first baroclinic Rossby deformation radius) and are mainly driven by mesoscale-induced local advection. An example can be found in Figure 4a, marked within the red boxes. As shown in the figure, in the deterministic forecast, although it can accurately forecast the emergence of the 2020 Northeast Pacific MHWs, many small structures of MHWs are ignored. However, enhanced by probabilistic model-based forecast ensembles, the small-scale MHW structures in the red boxes can be predicted with accurate intensity and shape. To further illustrate this multi-scale forecast issue, we also conduct a more detailed analysis based on the power spectrum of SSTA ground truth and forecast at different lead times in Figure 4d. Generally speaking, due to the overly smoothing issues in AI methods, our deterministic model shows a weaker SSTA spectrum at the mesoscale regime (corresponding to wave number about $10^{-4}m^{-1}$). At a shorter time scale (1 day), the improvement of the probabilistic model is mainly reflected in the strengthening of the mesoscale MHW structure. As the forecast lead time marches on, our probabilistic model is more inclined to improve MHWs forecast at a larger scale. In all, our probabilistic model can improve the model's ability to predict the structure of small-scale MHWs, which shows potential applications in regional forecasting.

\subsection{The Explainability of Our Model: Viewpoints from the Governing Equations}
In this section, we are going to focus on the explainability of our framework. In other words, we want to determine the key factors that drive the small-scale variations (10-day) of MHWs through the signed and unsigned contribution map described in section 2.4. By averaging equations (\ref{importance_ns}) and (\ref{importance_s}) over the decaying and intensifying phase of MHWs, we can quantify the role of each input variables in the life cycle of MHWs, as illustrated in Figure 5.\par

First we examine the contribution map of each input variable on the SSTA variations during intensifying/decaying phase of MHWs (i.e., the output of 
 $\phi^{(2)}_{\theta} \circ \phi^{(1)}_{\theta}$). As shown in Figure 5a, the contribution of surface wind speed is the largest among each variable, which leads to a maximum about $0.3$ K deviation of SSTA in both the intensifying and decaying phases. The surface temperature follows the wind speed and the contribution of SSTA is the smallest. The unsigned contribution of atmosphere and ocean variables is mainly concentrated in midlatitude and subarctic regions (e.g., Northeast Pacific and Western Boundary Currents). When considering the signed contribution, removing surface wind speed will generally lead to an underestimate in both intensifying and decaying phase of MHWs. This further emphasize the wind speed as a important predictor in the most extreme MHWs events.\par

When considering the contribution of inputs on each flux given by coupler (i.e. the output of $\phi^{(1)}_{\theta}$, Figure 5b), the contribution map still proves the surface wind speed as the key driving factor. More specifically, the contribution of surface wind on latent heat flux is the largest, where $\Delta C^{ns}_{j}$ is mainly confined within the subtropical regions, with peak values are found in the East Pacific and Western Australia. The influence on short-wave radiation is also relatively considerable, followed by sensible heat flux, and long-wave radiation is the least affected. In addition, large response of heat flux can be found along Western Boundary Currents, such as the Gulf Stream, Kuroshio Extension, and Agulhas System. Along these regions, eddy activities can feedback to the atmosphere, which will further maintain or strengthen the MHWs through air-sea heat flux \cite{holbrook2019global}. \par

However, as mentioned above, the spatial distribution of input variables on SSTA and air-sea heat flux is quite different, especially in the Northern Hemisphere (including the well-known Northeast Pacific MHWs). According to (\ref{heat budget}), the relatively deep mixed layer in the midlatitudes should reduce the response of MHWs to the same heat flux change. This contradiction indicates that the small-scale variations of MHWs in the midlatitude ocean is more sensitive to the changes in the heat flux. We hypothesize that it may be due to the intricate subsurface effect that drives the amplifying of MHWs in these regions. For example, the subsurface entrainment process is believed to drive the Central North Pacific MHWs, while the anomalous mixed layer depth can contribute to Northeast Pacific MHWs \cite{nishihira2024record,shi2022role}. In summary, in our framework, the surface wind speed is the key factor of the predictability of MHWs, which mainly controls the development of MHWs by affecting surface latent heat flux.

\section{Conclusions}

In this study, by designing a data-driven framework, we evaluate the predictability of small-scale (10-day) variations of global MHWs, which partly address the limitation of existing methods in forecasting extreme MHWs. We believe that this technique can be directly applied to other AI-based weather forecasts, thereby improving the model's ability to predict extreme events including heatwaves, extreme precipitation, and drought \cite{wu2024beamvq}. What's more, Through XAI methods, we draw the conclusion that wind forcing is the key factor that drives the small-scale variation of MHWs through impacting the air-sea latent heat flux. This understanding provides a foundational framework for future data-driven operational forecasts and research into the driving factors of MHWs. \par

However, there are still some limitations of our study, which can be carried out in the future. First, as shown in Figure 2a, The role of atmospheric forcing is crucial, especially in long-term MHW predictions. Improving the accuracy of subseasonal weather forecasts could significantly enhance the performance of our framework. Second, in our current framework, only surface data is used as input. Although our framework is able to parameterize some of these subsurface effects, by integrating the observation and reanalysis data from the subsurface layers, it is expected to further improve the explainability and forecast ability of our model. Third, it should be pointed out that the use of more physical variables based on the governing equations (\ref{heat budget}) can theoretically further enhance our forecasting. However, in order to ensure that our model is operational and lightweight as much as possible, we only used 8 intermediate variables (4 basic variables and 4 flux) in this study. Future work could revolve around the impact of various physical variables on MHW forecasts. Nevertheless, considering the scale-dependent nature of MHWs, it would be meaningful to investigate the predictability of the small-scale MHWs, especially with the advent of the Surface Water and Ocean Topography (SWOT) programs \cite{fu2024surface}.\par

%Third, it should be pointed out that the use of more physical variables based on the governing equations (\ref{heat budget}) can theoretically further enhance our forecasting. However, in order to ensure that our model is operational and lightweight as much as possible, we only used 8 intermediate variables (4 basic variables and 4 flux) in this study. Future work could revolve around the impact of various physical variables on MHW forecasts.

In summary, while our model offers a robust framework for MHW forecasting, addressing these issues and expanding the scope of our research will further refine our understanding and prediction capabilities of marine heatwaves, contributing to both scientific knowledge and operational forecasting.

%%%%%%%%%%%%%%%%%%%%%%%%%%%%%%%%%%%%%%%%%%%%%%%
%
% DATA SECTION and ACKNOWLEDGMENTS
%
%%%%%%%%%%%%%%%%%%%%%%%%%%%%%%%%%%%%%%%%%%%%%%%

\section*{Open Research}
The ERA5 reanalysis dataset can be downloaded from \url{https://cds.climate.copernicus.eu/datasets/reanalysis-era5-pressure-levels}. The reanalysis dataset of SST can be found at \url{https://data.marine.copernicus.eu/product/GLOBAL_MULTIYEAR_PHY_001_030} and \url{https://data.marine.copernicus.eu/product/GLOBAL_MULTIYEAR_PHY_ENS_001_031}. The GOFS 3.1 forecast results can be acquired at \url{https://www.hycom.org/dataserver/gofs-3pt1/analysis}. For the source code to reproduce our study, we have made the source code available online \cite{shu_2024_14189229}.

\acknowledgments
This work was supported by the National Natural Science Foundation of China (42125503, 42430602). Thanks to the National Key Scientific and Technological Infrastructure project “Earth System Science Numerical Simulator Facility” and the Global Water Cycle Observatory (No. 183311KYSB20200015) for providing computational resources and data support.

%%%%%%%%%%%%%%%%%%%%%%%%%%%%%%%%%%%%%%%%%%%%%%%
% REFERENCES and BIBLIOGRAPHY
%
% \bibliography{<name of your .bib file>} don't specify the file extension
% don't specify bibliographystyle
%
%%%%%%%%%%%%%%%%%%%%%%%%%%%%%%%%%%%%%%%%%%%%%%%

\bibliography{main_text}
\newpage
\begin{figure}
\noindent\includegraphics[width=\textwidth]{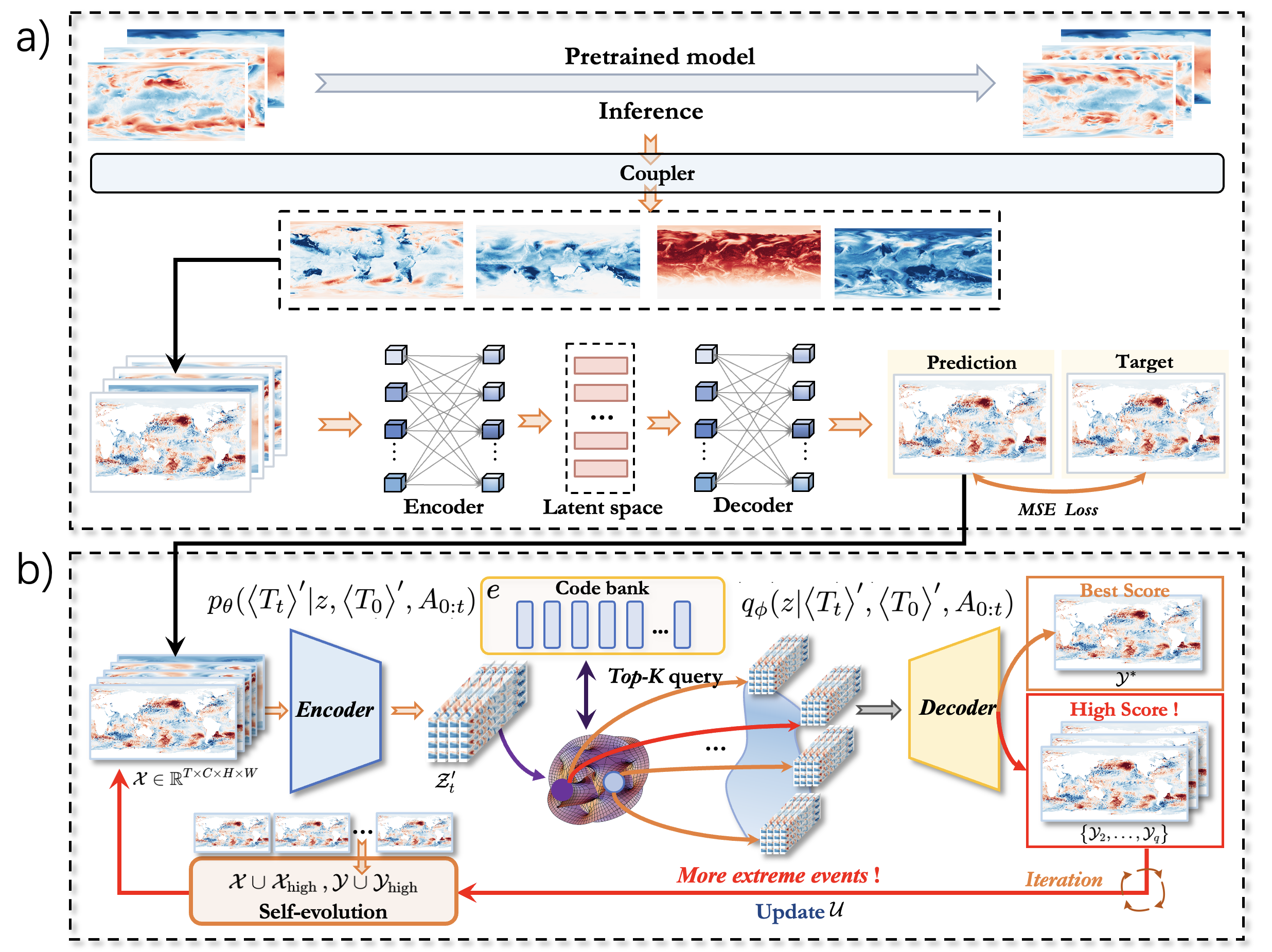}
\caption{The overall framework of our global MHWs forecast framework. \textbf{a)} The deterministic forecast.  \textbf{b)} The probabilistic part designed to improve extreme MHWs' prediction.}
\label{Figure 1.}
\end{figure}

\begin{figure}
\noindent\includegraphics[width=\textwidth]{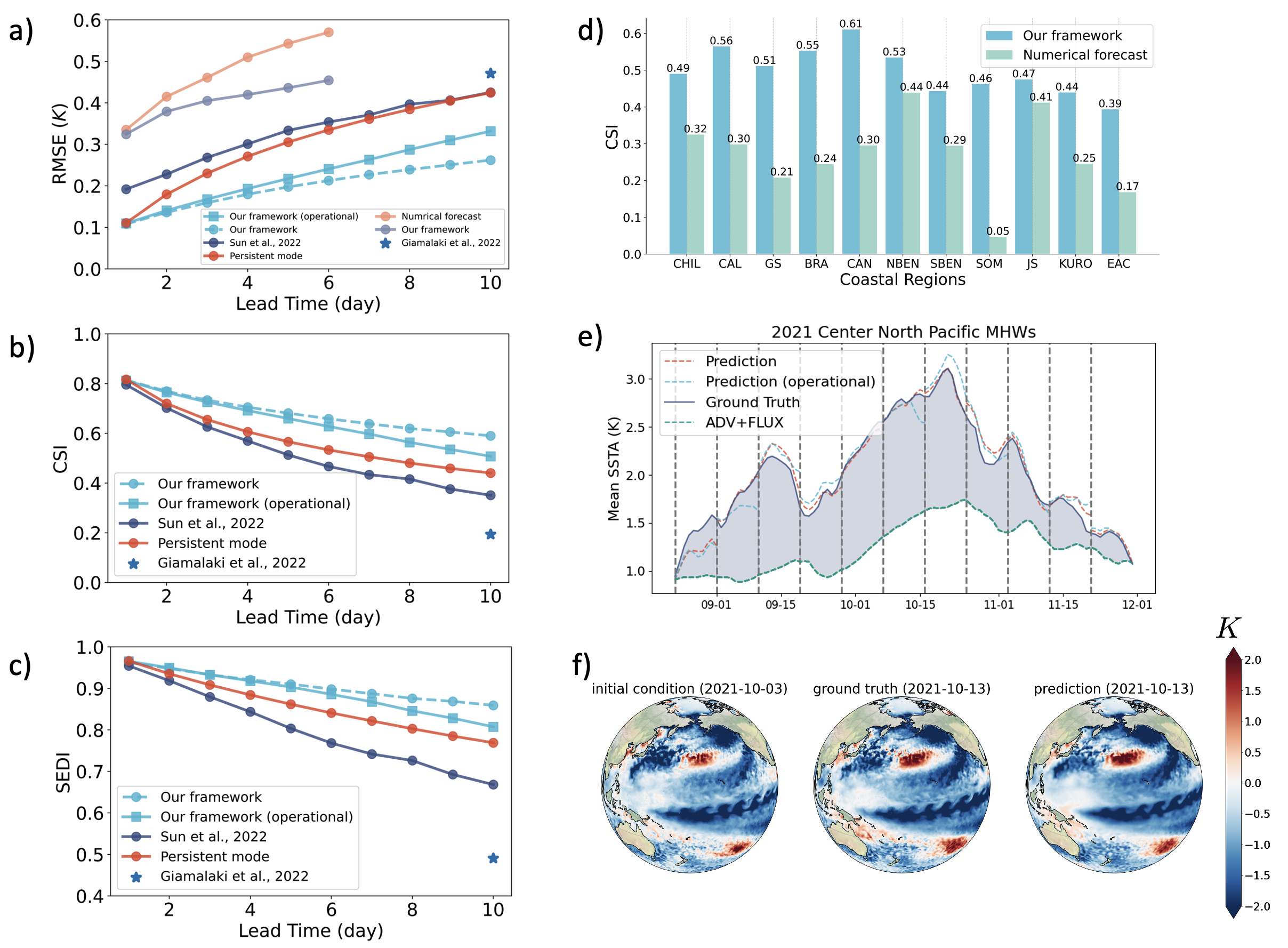}
\caption{The overall performance of our global MHWs forecast model. \textbf{a)-c)} The RMSE, CSI, SEDI of our framework comparing to existing subseasonal MHWs forecast at a 10-day lead. \textbf{d)} The CSI score around global coastal regions. \textbf{e)} The 10-day roll-out forecast of our framework initialized on 2021-08-22. \textbf{f)} The forecast result (SSTA) of our framework initialized on 2021-10-03. It is important to highlight that we deduct 90th percentile of the original SSTA data, hence the red part of this image indicates the MHW event (if it also lasts longer than 5 days).}
\label{Figure 2.}
\end{figure}

\begin{figure}
\noindent\includegraphics[width=\textwidth]{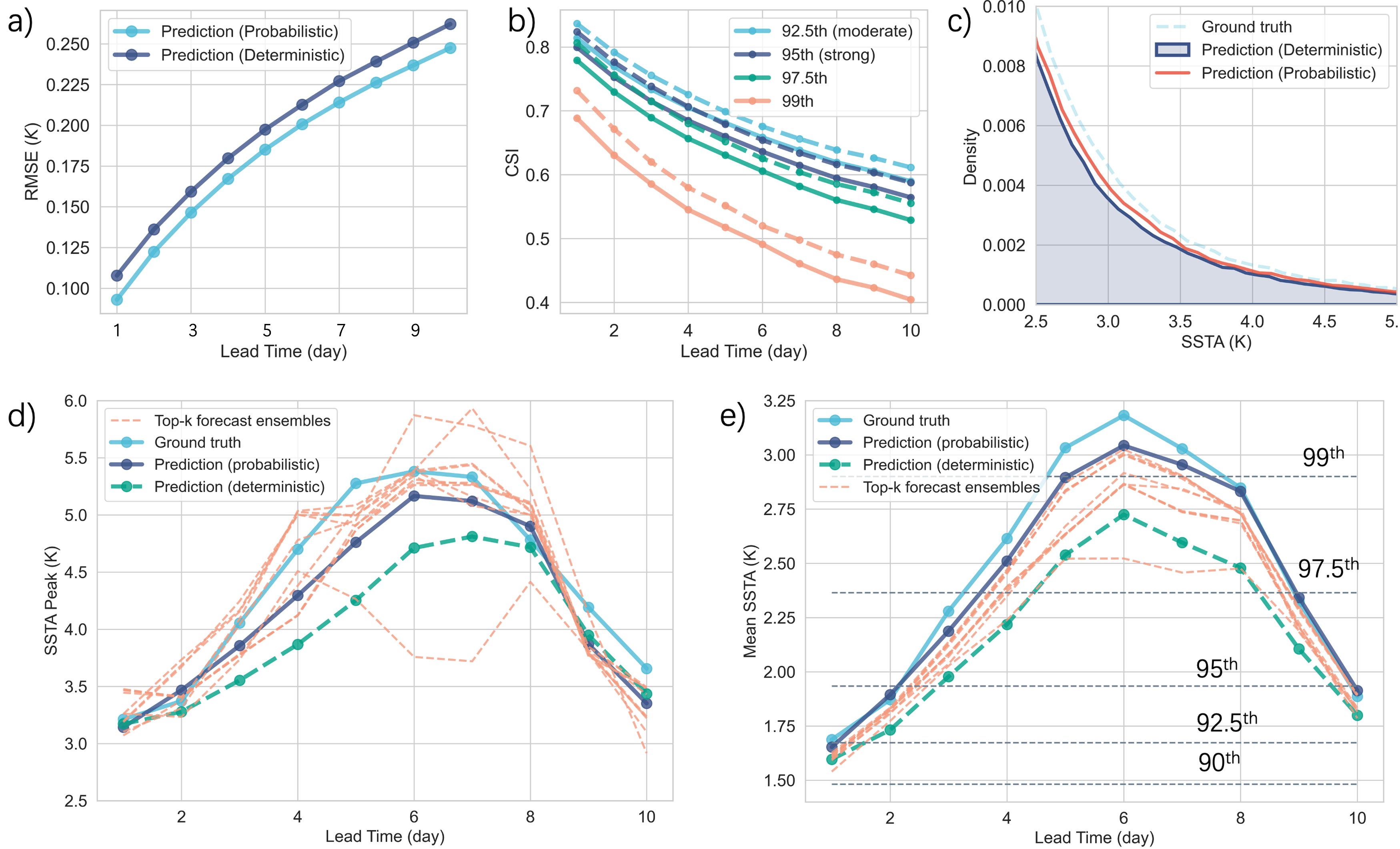}
\caption{A comparison of our deterministic and probabilistic forecasts. \textbf{a)-b)} The RMSE and CSI of our deterministic and probabilistic forecasts. \textbf{c)} The probability density function (PDF) of our deterministic and probabilistic forecast results. \textbf{d)-e)} A 10-day forecast of 2020 CNP-MHW initialized on 2020-08-14.}
\label{Figure 3.}
\end{figure}

\begin{figure}
\noindent\includegraphics[width=\textwidth]{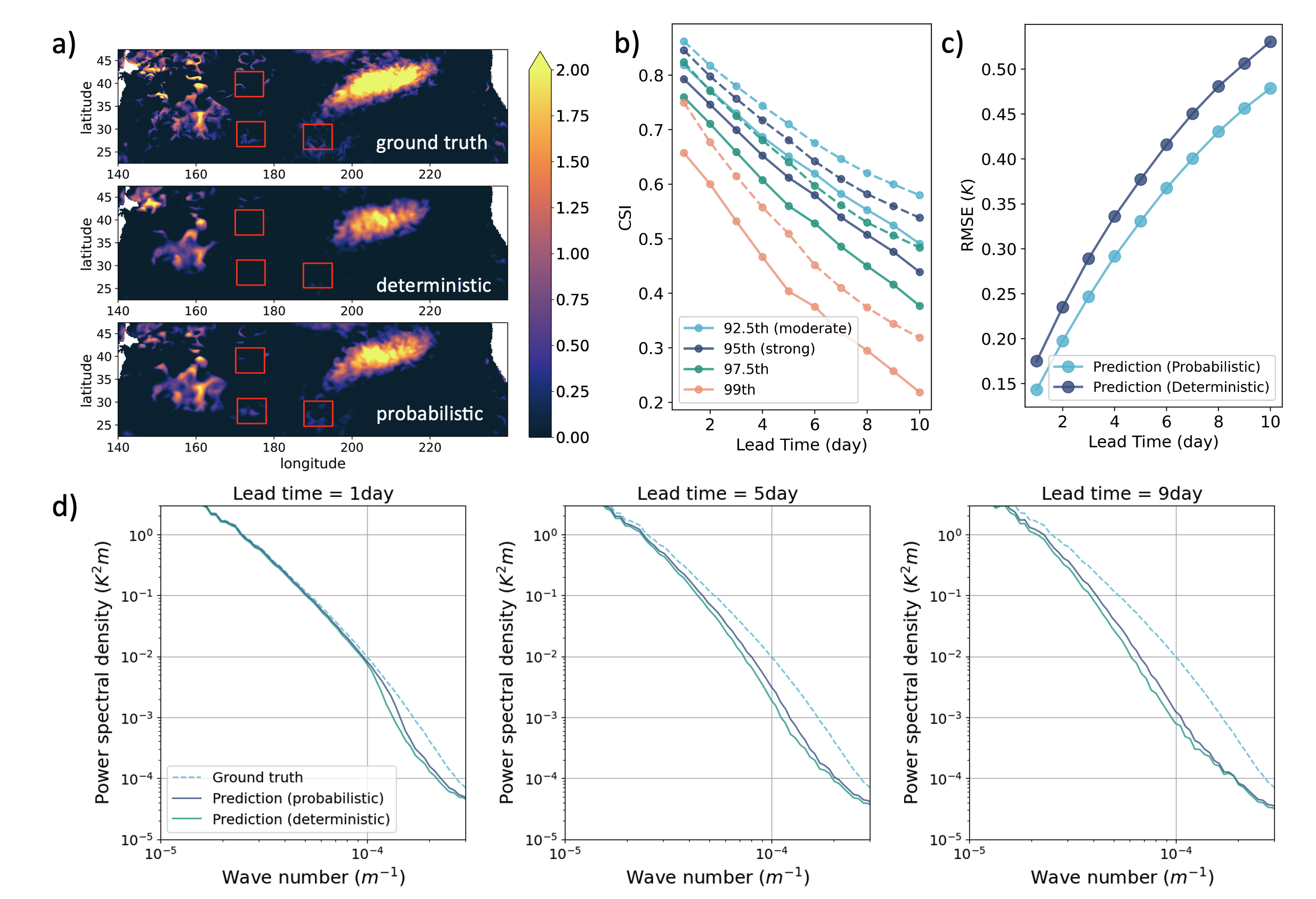}
\caption{The performance of our model in the high-resolution regional forecast (North Pacific). \textbf{a)} The 10-day forecast result of MHWs on 2020-07-17. The purple to orange parts of the image represent MHW events and the shade of color represents its intensity. \textbf{b)-c)} A comparison of our deterministic and probabilistic forecasts, similiar to Figure 3a-b. \textbf{d)} The power spectral of SSTA forecast at different lead time.}
\label{Figure 4.}
\end{figure}

\begin{figure}
\noindent\includegraphics[width=\textwidth]{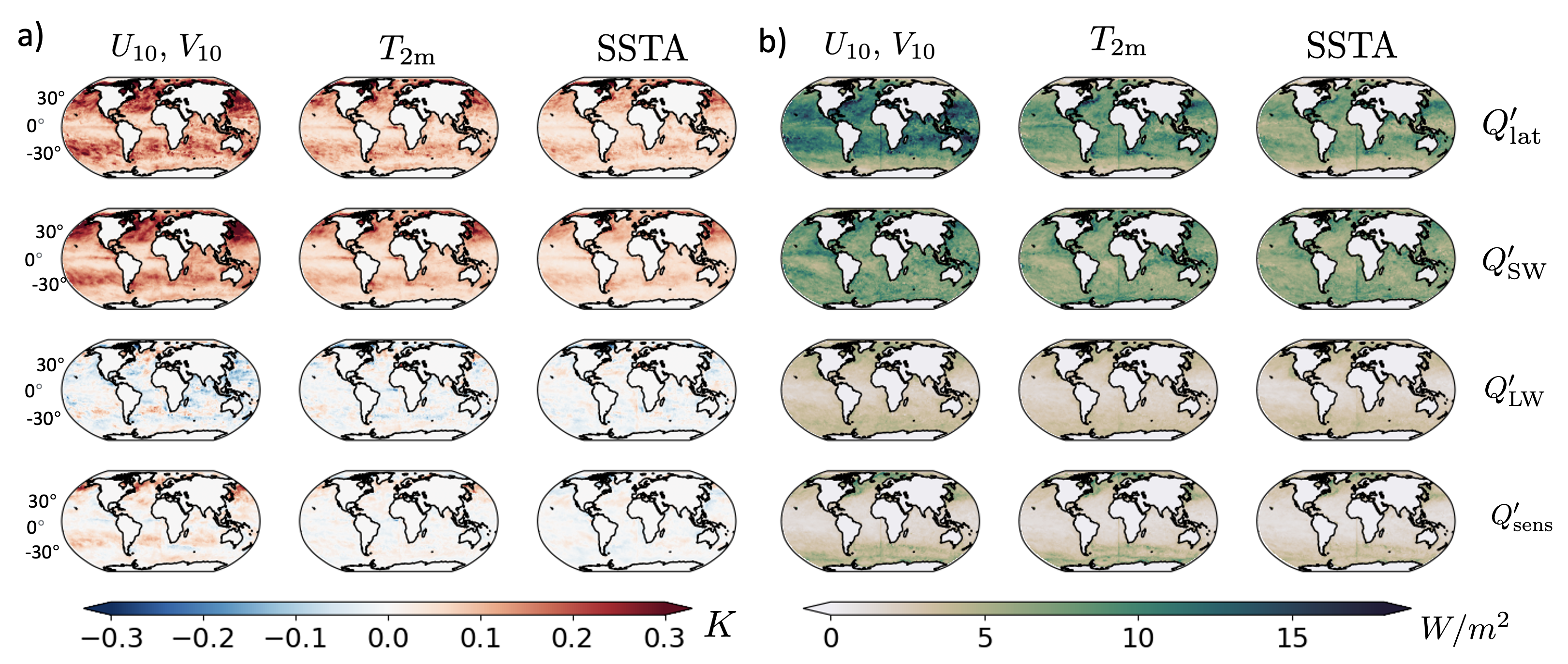}
\caption{The explainability of our model. \textbf{a)} The contribution map of each input variables on the SSTA during intensifying/decaying phase of MHWs. From top to the bottom: unsigned map during intensifying phase and decaying phase, signed map during intensifying and decaying phase.  b) The unsigned contribution map of each input variables on the coupler's outputs during the intensifying phase of MHWs.}
\label{Figure 5.}
\end{figure}

\end{document}